\documentclass[12pt]{article}
\usepackage{epsf}
\usepackage{graphicx}
\usepackage{a4}
\usepackage{amsmath}
\usepackage{amssymb}
\usepackage{cite}		
\usepackage{color}
\usepackage{tikz}
\usepackage{pgfplots}
\usepackage{empheq} 

\usepackage[colorlinks=true, linkcolor=blue, citecolor=blue, urlcolor=blue]{hyperref}
\usepackage{hyperref}
\usepackage{pdfcomment}
\usepackage{tikz}
\usetikzlibrary{arrows.meta}

\def\hybrid{\topmargin 0pt
        \oddsidemargin 0pt
        \headheight 0pt \headsep 0pt
        \textwidth 6.25in       
        \textheight 9.in       
        \marginparwidth .875in
        \parskip 5pt plus 1pt   \jot = 1.5ex}

\catcode`\@=11
\def\marginnote#1{}

\newcount\hour
\newcount\minute
\newtoks\amorpm
\hour=\time\divide\hour by60
\minute=\time{\multiply\hour by60 \global\advance\minute by-\hour}
\edef\standardtime{{\ifnum\hour<12 \global\amorpm={am}%
        \else\global\amorpm={pm}\advance\hour by-12 \fi
        \ifnum\hour=0 \hour=12 \fi
        \number\hour:\ifnum\minute<10 0\fi\number\minute\the\amorpm}}
\edef\militarytime{\number\hour:\ifnum\minute<10 0\fi\number\minute}

\def\draftlabel#1{{\@bsphack\if@filesw {\let\thepage\relax
   \xdef\@gtempa{\write\@auxout{\string
      \newlabel{#1}{{\@currentlabel}{\thepage}}}}}\@gtempa
   \if@nobreak \ifvmode\nobreak\fi\fi\fi\@esphack}
        \gdef\@eqnlabel{#1}}
\def\@eqnlabel{}
\def\@vacuum{}
\def\draftmarginnote#1{\marginpar{\raggedright\scriptsize\tt#1}}

\def\draft{\oddsidemargin -.5truein
        \def\@oddfoot{\sl preliminary draft \hfil
        \rm\thepage\hfil\sl\today\quad\militarytime}
        \let\@evenfoot\@oddfoot \overfullrule 3pt
        \let\label=\draftlabel
        \let\marginnote=\draftmarginnote
   \def\@eqnnum{(\theequation)\rlap{\kern\marginparsep\tt\@eqnlabel}%
\global\let\@eqnlabel\@vacuum}  }

\def\numberbysection{\@addtoreset{equation}{section}
        \def\theequation{\thesection.\arabic{equation}}}

\def\titlepage{\@restonecolfalse\if@twocolumn\@restonecoltrue\onecolumn
     \else \newpage \fi \thispagestyle{empty}\c@page\z@
        \def\thefootnote{\fnsymbol{footnote}}
	\setcounter{page}{0} }
\def\endtitlepage{\if@restonecol\twocolumn \else  \fi
        \def\thefootnote{\arabic{footnote}}
        \setcounter{footnote}{0}}  
\definecolor{c1}{rgb}{1, 0, 0}
\definecolor{c2}{rgb}{0, 1, 0}
\definecolor{c3}{rgb}{0, 0, 1}
\definecolor{c4}{rgb}{1, 0, 1}
\definecolor{c5}{rgb}{0, 1, 1}

\catcode`@=12
\relax

\def\ie{\hbox{\it i.e.}}

\def\beq{\begin{equation}}
\def\eeq{\end{equation}}
\def\bea{\begin{eqnarray}}
\def\eea{\end{eqnarray}}
\def\EQ{\begin{equation}}
\def\EN{\end{equation}}

\relax
\numberbysection
\hybrid

\usepackage{setspace}

\begin{document}

\begin{center}
{\large\bf Phase Diagram and Critical Behaviour of the}\\[4pt]
{\large\bf  Two-Dimensional Potts Model with}\\ [4pt]
{\large\bf   Long-Range Quenched Disorder}\\[6pt]
{\bf Rodolfo Rocha, Leticia F. Cugliandolo, Marco\ Picco}\\
  Sorbonne Universit\'e \& CNRS, UMR 7589, LPTHE, F-75005, Paris, France\\
    e-mail: {\tt rocha,leticia,picco@lpthe.jussieu.fr} \\
\end{center}

\centerline{(Dated: \today)}
\vskip .2in
\centerline{\bf ABSTRACT}
\begin{quotation}
We study the phase diagram and critical properties of the $q = 3$ and $q = 8$ Potts models with spatially correlated disorder governed 
by a power-law decay with exponent $a$. Building on the phase diagram proposed by Chippari \textit{et al.}~\cite{CPS1}, 
where a transition from finite-disorder to infinite-disorder fixed points was identified as a function of $a$, we refine this picture by 
using wrapping probabilities of Fortuin--Kasteleyn clusters. Furthermore, using magnetic observables, we clarify the physical origin 
of the double-peak structure in the magnetic susceptibility and establish the validity of the hyper-scaling relation across all investigated regimes.
\end{quotation}

\begingroup
\setstretch{0.2} 
\tableofcontents
\endgroup

\section{Introduction}
\label{sec:Introduction}

More than forty years ago, Weinrib and Halperin~\cite{WH,W} investigated critical phase transitions in systems with long-range correlated
quenched disorder.
For disorder correlations with a slow algebraic decay, 
$x^{-a}$ with $a<d$, these authors derived an extended Harris criterion:  disorder is relevant if
$\nu<2/a$, where $\nu$ is the pure system's thermal critical exponent. 
If $a>d$, the randomness behaves as if it were short-ranged, and one recovers the standard Harris criterion.
For relevant disorder, a new fixed point characterised by the exponent $\nu_{LR} =2/a$ was predicted. 
Only Gaussian disorder was  considered in this work.

In some recent studies~\cite{CPS1,CPS2,LPS}, a similar analysis of the two-dimensional $q$-state Potts models~\cite{Wu} with 
 long-range correlated disorder was considered. In these works, the disorder was 
introduced by coupling each bond of the lattice to the product of $n$ auxiliary Ising spins. 
By construction, this disorder is non-Gaussian, and this raises the question of the validity of the predictions of Weinrib and Halperin beyond the 
Gaussian case.

Using a highly performant construction of algebraically correlated disorder
which exploits the thermal fluctuations of independent auxiliary systems,
Chippari {\it et al.} were able to study the disordered two dimensional
Potts model numerically~\cite{CPS1}. These authors showed that, for $q=2$ and $q=3$, there exists a fixed point 
at finite disorder, that is to say a phase transition at finite temperature, when $a$ is not too small ($a \geq 1.0$)
but still smaller than the spatial dimension ($d=2$). 
In addition, they characterised the fixed points  corresponding to infinite disorder. 
By construction, these infinite-disorder fixed points are identical for Potts models with an arbitrary number of states. 
For small values of $a$, \ie\ $a \leq 0.5$, the disorder drives the system towards these infinite-disorder fixed points. 
In the region $ 0.5 < a < 1.0$, it was not possible to distinguish between a finite disorder fixed point 
or a flow towards a fixed point at infinite disorder. 

One of the main advantages of the construction of correlated disorder
just described, as compared to the Gaussian disorder one (or to Chatelain's construction~\cite{C}), is that it also allowed
Chippari {\it et al.} 
to implement a perturbative renormalisation-group calculation based on conformal perturbation theory through a double expansion 
in the parameters $2-a$ and $q-2$~\cite{CPS2}. 
This approach enabled these authors to derive predictions for the thermal critical behaviour, including deviations from the Weinrib--Halperin 
prediction for the correlation-length exponent, $\nu_{LR}=2/a$, arising from the non-Gaussian nature of the disorder.

In the present work, we present
numerical results for $q=3$ which give further support to some of the predictions in~\cite{CPS2}.
We also extend the analysis to the $q=8$ case, where the introduction of both uncorrelated~\cite{P,CJ,CB} or correlated \cite{C} 
disorder soften the first-order phase transition into a second-order one.

In the following, we will compute the wrapping probability of Fortuin--Kasteleyn clusters~\cite{FK} 
on the torus and use it as a probe of the underlying fixed points. 
In the pure case, this quantity matches the known exact value, while in the strong-disorder regime it approaches values consistent with correlated percolation. By analysing finite-size scaling and collapse properties of suitably rescaled observables, we determine the effective disorder strength associated with critical behaviour for different values of $a$.

Our results confirm that for $a=0.5$ the system flows towards an infinite-disorder fixed point, with no crossing of wrapping probabilities and a robust scaling collapse consistent with percolation-like behaviour. For $a=1$, we identify a finite-disorder attractive fixed point, while for $a=0.75$ we observe a crossover regime with a broad but finite range of effective critical disorder, suggesting proximity to a transition between the two regimes.

We further analyse the magnetic susceptibility and show that for $a=0.5$ a double-peak structure emerges, which was previously interpreted as a possible Griffiths phase~\cite{C,C3}. However, finite-size scaling of the peak positions indicates that this structure is not associated with a true extended critical phase. Instead, it is compatible with the presence of a single critical point with strong finite-size effects, particularly for intermediate values of $a$.

Finally, we provide a physical interpretation of the infinite-disorder limit in terms of a two-scale activation process of strong and weak bonds. In this regime, the model reduces to bond percolation at criticality, and the characteristic temperature scales associated with bond activation separate with no bound.

The structure of the paper is the following. In Sec.~\ref{sec:Potts-model} we recall the definition of the $q$-state Potts model with correlated disorder.
We present our results for the $q=3$ case in Sec.~\ref{sec:q3Potts} and for the $q=8$ case in Sec.~\ref{sec:q8Potts}. 
In Sec.~\ref{sec:hyperscaling} we perform a hyper-scaling analysis. 
Finally, we conclude in Sec.~\ref{sec:conclusions}.

\section{Potts model with correlated disorder}
\label{sec:Potts-model}

The $q$-state Potts model is a generalisation of the Ising model in which each lattice site holds a spin that can take 
one of $q$ discrete values. In the presence of quenched bond disorder, the Hamiltonian of the system is 
\begin{equation}
\mathcal{H} = -\sum_{\langle i,j \rangle} J_{ij} \, \delta_{\sigma_i, \sigma_j} 
\; , 
\end{equation}
where $\sigma_i \in \{1, 2, \dots, q\}$ represents the spin state at the lattice site $i$, 
 $\langle i,j \rangle$ denotes a sum over nearest-neighbour pairs on the lattice,
  $\delta_{\sigma_i, \sigma_j}$ is the Kronecker delta which equals $1$ if $\sigma_i = \sigma_j$ and $0$ otherwise, 
 and $J_{ij}$ is the disordered (quenched) exchange coupling between sites $i$ and $j$.

In the present work, the disorder is built as follows. We simulate $n$ independent copies of the 
Ising model, $s^{(1)}, \cdots, s^{(n)}$,  on the same lattice as the one used for the Potts model. 
All these Ising models are simulated at their critical point.
After letting them equilibrate, we compute, for each point $i$ on the lattice, the variable 
\begin{align}
\tau_{i} = \prod_{k=1}^n  s_i^{(k)} \; , 
\end{align}
which takes values $\pm 1$.
In this work, we consider only two-dimensional square lattices. At each site $i$, the two bonds of the Potts model, $J_i^x$ and $J_i^y$, connecting $i$ to its right 
and upper nearest neighbours, respectively, take one of the two values, $J_1$ or $J_2$, 
of a bimodal disorder distribution according to the value taken by $\tau_i$: 
\begin{align}
J^x_i= J^y_i = \frac{J_1 + J_2}{2} + \tau_i \,  \frac{J_1 - J_2}{2}  
\; . 
\label{eq:Jx}
\end{align}
By construction, these bonds satisfy 
 \begin{equation}
 \langle J_i J_j \rangle \simeq |{\boldsymbol x}_i-{\boldsymbol x}_j|^{-a}
 \end{equation}
 with ${\boldsymbol x}_i$ and ${\boldsymbol x}_j$ the positions of sites $i$ and $j$, respectively, 
 $|{\boldsymbol x}_i-{\boldsymbol x}_j|$ the distance between these two sites, and 
 \begin{equation}
 a = \frac{n}{4}
 \; . 
 \end{equation} 
Since $\tau_i = \pm 1$, $J_1$ and $J_2$ are the two values each bond can take with equal probability 
and the parameter that controls the strength of disorder is $r=J_1/J_2$. Indeed, for $r=1$ one recovers the clean model without disorder,
while the  limit $r \rightarrow \infty$ (or equivalently $r \rightarrow 0$) corresponds to percolation. 
It is convenient to introduce randomness 
with this bimodal distribution because, for any value of the parameter $r$, if $J_1$ and $J_2 = J_1/r$ are chosen such that the relation
\begin{align}
\label{kd}
(e^{J_1} -1 )( e^{J_2}-1) = q 
\end{align}
holds, then it corresponds to a critical point separating a ferromagnetic phase from a paramagnetic one \cite{KD}. All the line $J_2 = J_1/r$
with $r \in [1,\infty[$ is critical 
as a function of a thermal perturbation. Note that, with these conventions, we implicitly assume that $\beta=1$ at the critical points; otherwise stated, we 
measure $J_1$ and $J_2$ in units of temperature. 
Thus, for a given configuration of the $n$ copies of the Ising model, one can generate a corresponding configuration of correlated bonds. The copies of the Ising model are then updated to produce a new bond configuration, and the procedure is repeated.

A similar construction based on the Ashkin-Teller model was previously introduced by Chatelain~\cite{C}, although it is limited to small values of $a$.
With the construction above, we do not face such limitation.

We will focus on the two dimensional square geometry. In the pure ferromagnetic case, $J_{ij} = J$ for all nearest neighbour pairs $ij$, equivalently $r=1$, 
 the two-dimensional Potts model has a phase transition  between a low temperature ferromagnetic phase and 
 a high temperature paramagnetic one at $(\beta J)_c = \ln (1 + \sqrt{q})$.
 The transition is second order  for $q\leq 4$ and first order for $q>4$~\cite{Wu}. This is modified by 
 randomness \cite{Ludwig,Ludwig2,DPP1,DPP,P,CJ,CB} especially if long-range correlated \cite{CPS1}, as discussed below. 
  
 In the simulations we use a system with linear size $L$ and periodic boundary conditions, that is to say, 
 we work on a torus.
 
\section{$q=3$ Potts model}
\label{sec:q3Potts}

We will first study the simple three colour $q=3$ case and we consider different values of $a$, the parameter that 
controls the range of the disorder correlations. We fix $T=1$ and use 
$r=J_2/J_1$ as the control parameter. 

\subsection{Phase diagram and critical points}

In Ref.~\cite{CPS1}, a sketch of the phase diagram was obtained for this model. In particular, it was shown that there is a finite-disorder 
fixed point for $a \geq 1$ (a finite temperature phase transition), while for $a \leq 0.5$ the fixed 
point is at infinite disorder. For $a = 0.75$, the situation remained unclear.

We aim to make this picture more precise by obtaining a better determination of the disorder strength at the critical points. 
In \cite{CPS1}, the disorder strength associated with the fixed points was only determined rather crudely, by studying 
the effective magnetic exponent as a function of $L$ and checking its stability as the system size increased.

Here, we compute the wrapping probability, $W(r,L)$, which is the probability that at least one Fortuin--Kasteleyn cluster wraps around the torus. 
For the pure $q=3$ model, corresponding to $r=1$, this quantity is known exactly~\cite{Arguin}: 
\begin{equation}
W_3 = 0.8137
\; , 
\label{eq:W3}
\end{equation}
where the subscript 3 specifies the 
number of colours, $q=3$. In the limit of strong disorder, it approaches 
the value for correlated percolation. For uncorrelated percolation, this value was determined to be $W_1^{u} = 0.6905$ \cite{Pinson} 
(the subscript 1 indicates $q\to 1$ for percolation). In the case of 
correlated percolation, it was shown in \cite{CPS1} that this quantity depends on $a$. For $a = 0.25$, it was found that $W_1^{a=0.25} = 0.5159$, 
and this result was verified numerically.

Numerical measurements for other values of $a$ were not reported in \cite{CPS1}. By reanalysing 
the data produced in that work, we obtain
\begin{align}
W_1^{a=0.5} &= 0.5340, & W_1^{a=0.75} &= 0.5580, & W_1^{a=1} &= 0.5915.
\label{eq:W1}
\end{align}

\begin{figure}[ht]
\centerline{
(a) \hspace{7cm} (b) \hspace{3.5cm}
}
\begin{center}
\includegraphics[width=7.75cm,height=6.5cm]{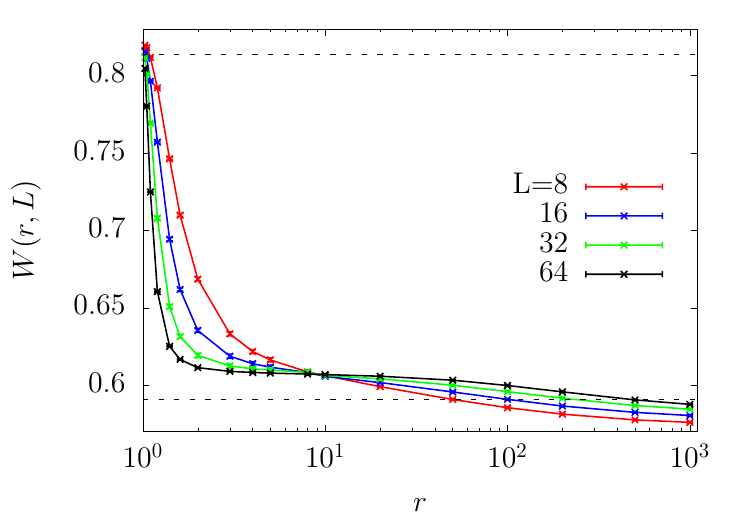}
\includegraphics[width=7.75cm,height=6.5cm]{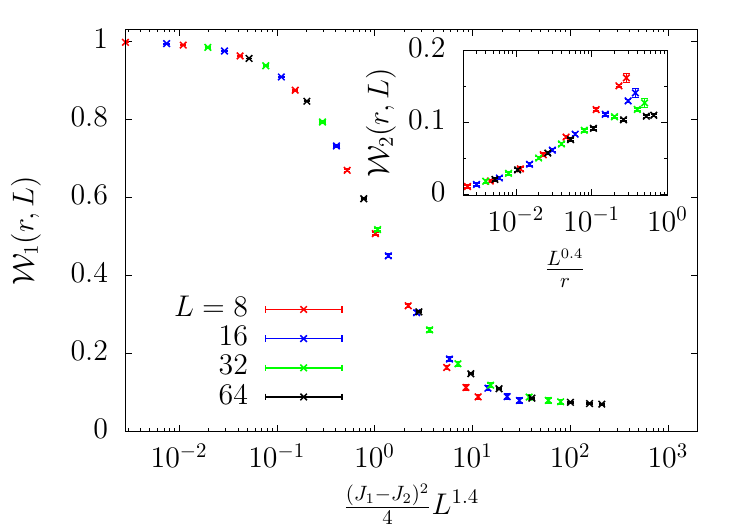}
\end{center}
\vspace{-0.5cm}
\caption{$q=3$ Potts model with $a=1$. 
(a) Wrapping probability $W(r,L)$ as a function of the parameter $r=J_2/J_1$.
The horizontal dashed lines are at $W_3=0.8137$ and $W_1^{a=1}=0.5915$.
(b) ${\mathcal W}_1(r,L)$ defined in Eq.~(\ref{DW1}) vs. $\frac{(J_1-J_2)^2}{4} L^{1.4}$  and 
${\mathcal W}_2(r,L)$ defined in Eq.~(\ref{DW2}) vs. $L^{0.4}/r$ in the inset.
See the text for details on these quantities. Different data points correspond to different system sizes given in the keys.
}
\label{FigQ3a1}
\end{figure}

We first consider the case $a=1$. In Fig.~\ref{FigQ3a1}(a), we show the wrapping probability as a function of 
the disorder strength $r$ for increasing linear sizes $L$. The wrapping probability interpolates between the values $W_3$ 
and $W_1^{a=1.0}$ (shown as dashed lines), with a crossing occurring at a disorder strength $r \simeq 10$ and a wrapping 
probability of approximately $0.61$.  Here and in the following, unless otherwise stated, each measurement is averaged over $10^6$ disorder realisations.

It is important to note that, since this fixed point is attractive, a precise determination of its location is not required. 
To verify that it is indeed an attractive fixed point, it is sufficient to show that the two other fixed points are 
non-attractive.\footnote{An alternative approach would be to determine a scaling law in the vicinity of the new fixed point. 
This is difficult because the deviation is expected to behave as $(r-r_c)L^{-|y|}$, where $r_c$ is the disorder strength at 
the new fixed point and $-|y|<0$. Close to the fixed point, this contribution is mixed with corrections due to irrelevant 
operators of the form $\alpha L^{-\omega}$.}

In the weak-disorder limit, one expects that $W(r,L)$ scales as a function of $\frac{(J_1-J_2)^2}{4} L^{y_d}$, with $y_d$ 
the scaling dimension of the disorder perturbation \cite{LPS}. 
$\frac{(J_1-J_2)^2}{4}$ corresponds to the second cumulant of the disorder, which will couple with the leading perturbation \cite{Ludwig,DPP}. 
If disorder is a relevant perturbation, then $y_d>0$. 
For the $q=3$ Potts model, the scaling dimension of the disorder perturbation is predicted to be \cite{CPS2}
\begin{align}
\label{defsd}
y_d = 2-a+0.4 \; .
\end{align}
Since finite-size corrections are present (as can be clearly seen in the small- and large-disorder limits of $W(r,L)$), 
it is more convenient to consider the quantities
\begin{align}
\label{DW1}
{\cal W}_1(r,L) = \frac{1}{W_3-W_1^{a=1}}
\left(
W_3\frac{W(r,L)}{W_3(L)} - W_1^{a=1}
\right)
\end{align}
and
\begin{align}
\label{DW2}
{\cal W}_2(r,L) = \frac{1}{W_3-W_1^{a=1}}
\left(
\frac{W(r,L)}{W_1^{a=1}(L)} - 1
\right).
\end{align}
$W_3$ and $W_1^{a=1}$ are equal to the theoretical values given in Eq.~(\ref{eq:W3}) and 
Eq.~(\ref{eq:W1}), respectively, while $W_3(L)$ and $W_1^{a=1}(L)$ are the ones measured numerically
for the corresponding size~$L$.

The first quantity interpolates between $1$ in the weak-disorder limit and $0$ in the strong-disorder limit. It is constructed 
so as to eliminate finite-size corrections in the weak-disorder regime and is therefore expected to be a simple function of 
$\frac{(J_1-J_2)^2}{4}L^{y_d}$ for not too strong disorder. 

The second quantity is constructed to vanish in the strong-disorder limit while taking finite-size corrections in this regime into account. 
Since the precise form of the scaling variable at infinite disorder is not known, we simply test the scaling variable $L^{y_p}/r$. 
Note that, although the exponent at infinite disorder is denoted by $y_p$, indicating its association with percolation, it actually corresponds to correlated percolation. 
Therefore, one expects $y_p$ to depend on the correlation exponent $a$. A positive value of $y_p$ indicates that the infinite-disorder fixed point is unstable.

Both quantities are shown in Fig.~\ref{FigQ3a1}(b). The analytical prediction Eq.~(\ref{defsd}) for $a=1$ is $y_d=1.4$, 
and we indeed observe an excellent scaling collapse for this value. This prediction was already tested in \cite{LPS}, 
where a similar analysis was performed using the magnetisation. In the strong-disorder regime, there is data collapse  for 
different system sizes when plotted as a function of $L^{y_p}/r$ with $y_p=0.4$. This then confirms that a finite-disorder fixed point exists.

\begin{figure}[ht]
\centerline{
(a) \hspace{7cm} (b) \hspace{3.5cm}
}
\begin{center}
\includegraphics[width=7.75cm,height=6.5cm]{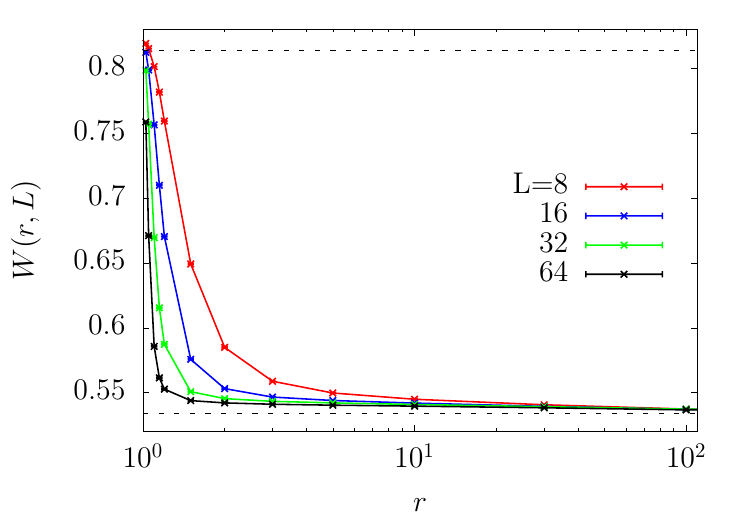}
\includegraphics[width=7.75cm,height=6.5cm]{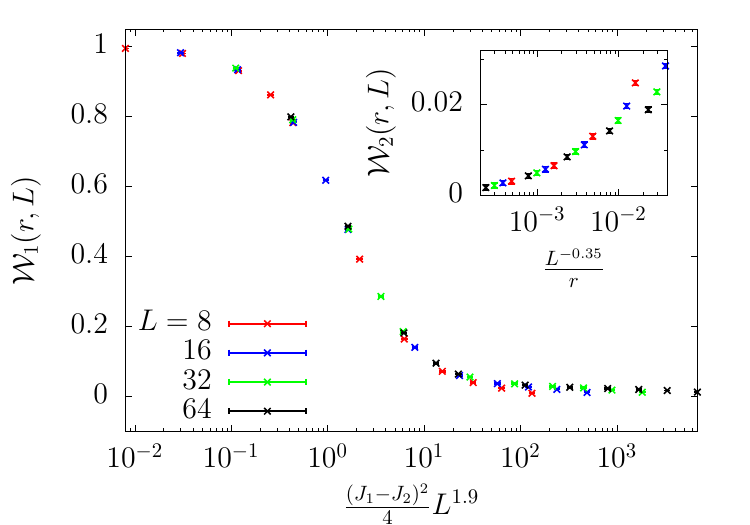}
\end{center}
\vspace{-0.5cm}
\caption{$q=3$ Potts model with $a=0.5$.
(a) Wrapping probability as a function of the disorder strength $r$. 
The horizontal dashed lines are at $W_3=0.8137$ and $W_1^{a=0.5}=0.5340$.
(b) ${\mathcal W}_1(r,L)$ vs. $\frac{(J_1-J_2)^2}{4} L^{1.9}$. The inset shows ${\mathcal W}_2(r,L)$ vs. $L^{-0.35}/r$. The linear sizes and colour code
is the same as in other figures and it is recall in the key.
}
\label{FigQ3a05}
\end{figure}

Next, we present analogous results for $a=0.5$ in Fig.~\ref{FigQ3a05}. The wrapping probability curves do not exhibit any crossing. 
In the right panel, we plot ${\cal W}_1(r,L)$ as a function of $\frac{(J_1-J_2)^2}{4} L^{y_d}$, with $y_d=1.9$ from Eq.~(\ref{defsd}). We obtain an 
excellent collapse of data over a large range of disorder strengths and system sizes. The inset shows ${\cal W}_1(r,L)$ as a function
of $L^{y_p}/r$ with $y_p=-0.35$. Note that the sign is negative now, so the strong-disorder limit point is attractive. 
This provides strong support for our previous claim that, for this value of $a$, the critical point is located at infinite disorder. 

\begin{figure}[!ht]
\centerline{
(a) \hspace{7cm} (b) \hspace{3.5cm}
}
\begin{center}
\includegraphics[width=7.75cm,height=6.5cm]{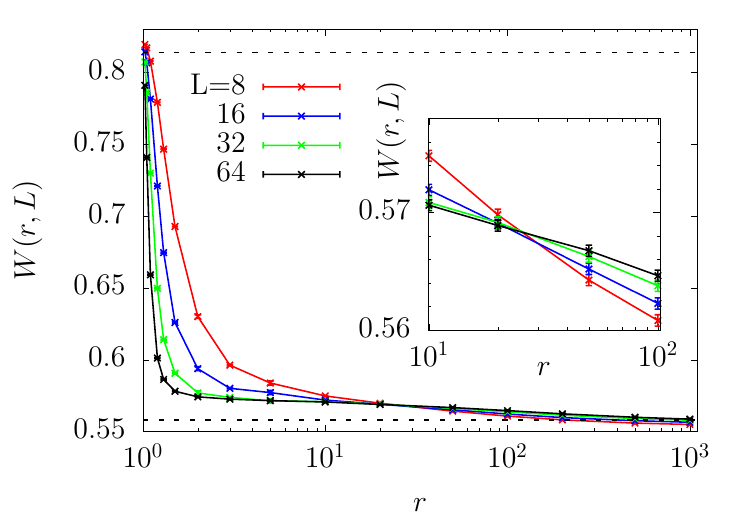}
\includegraphics[width=7.75cm,height=6.5cm]{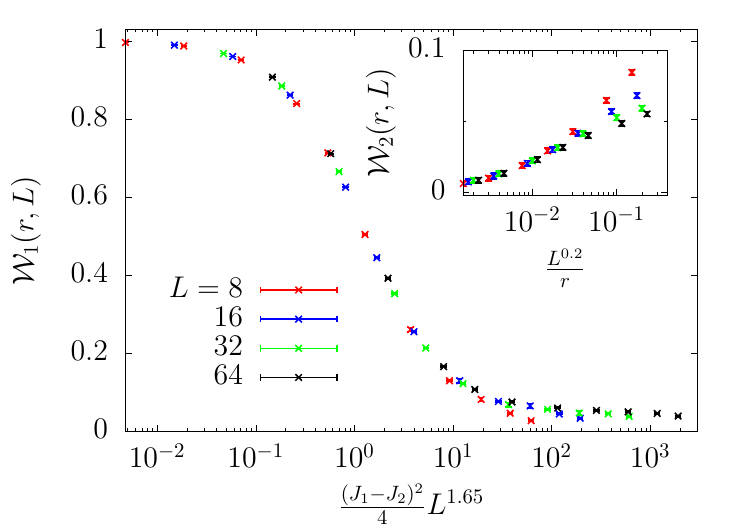}
\end{center}
\vspace{-0.5cm}
\caption{$q=3$ Potts model with $a=0.75$ and different linear system sizes given in
the keys.   (a) Wrapping probability as a function of the disorder strength $r$.
In the inset, a magnification around the crossing point. 
The horizontal dashed lines are at $W_3=0.8137$ and $W_1^{a=0.75}=0.5580$.
(b) ${\mathcal W}_1(r,L)$ vs. $\frac{(J_1-J_2)^2}{4} L^{1.65}$. The inset  shows ${\mathcal W}_2(r,L)$ vs. $L^{0.2}/r$.
}
\label{FigQ3a075}
\end{figure}

Finally, we consider the case $a=0.75$. In this case, we observe a crossing of the wrapping probability curves at a relatively 
large disorder strength, $r \simeq 20$, as shown in Fig.~\ref{FigQ3a075}(a). Since scaling corrections are present, 
the precise location of the crossing point is difficult to determine. This is where the quantities ${\cal W}_1(r,L)$ and ${\cal W}_2(r,L)$ 
introduced above become particularly useful.

In  Fig.~\ref{FigQ3a075}(b), we plot ${\cal W}_1(r,L)$ as a function of $\frac{(J_1-J_2)^2}{4} L^{y_d}$, 
with $y_d=1.65$ as predicted by Eq.~(\ref{defsd}). The data collapse is very good up to $r \simeq 10$. In the
inset, we plot ${\cal W}_2(r,L)$ as a function of $L^{y_p}/r$ and obtain a good collapse with $y_p=0.2$ 
in the large-disorder regime, extending down to approximately $r \simeq 50$.

These results indicate the existence of a finite-disorder fixed point, which lies in the range $10 < r_c < 50$. In the following, 
we therefore adopt the representative value $r=25$ for the case $a=0.75$.

\subsection{Magnetic susceptibility}

Next, we present the results of our simulations for the magnetic susceptibility in the three cases: 
(i) $a=0.5$ and $r=100$; (ii) $a=0.75$ and $r=25$; and (iii) $a=1.0$ and $r=10$.
We consider here the non-critical behaviour, and we thus reintroduce $\beta=1/T$. For a fixed disorder $r$ and $J_2=r J_1$
satisfying the relation (\ref{kd}), the bonds are $\beta J_1$ or $\beta J_2$. 

\begin{figure}[!h]
\centerline{
\hspace{-0.8cm} (a) \hspace{4.5cm} (b) \hspace{4.5cm} (c) \hspace{2cm}
}
\begin{center}
\includegraphics[width=5.2cm,height=4.3cm]{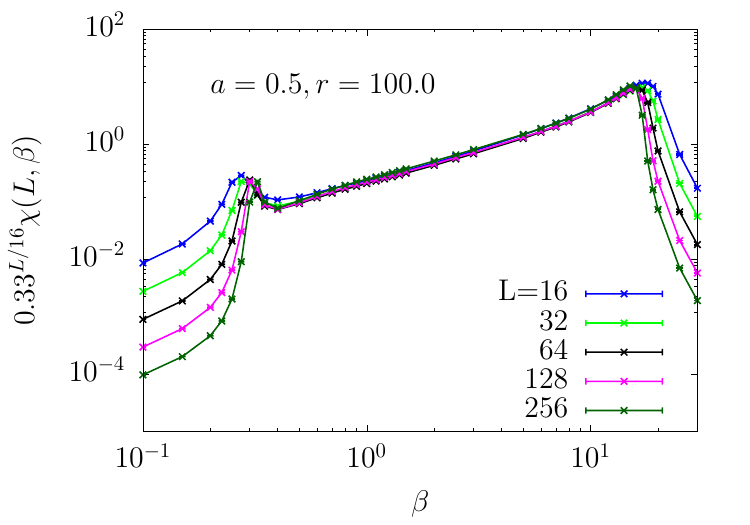}
\includegraphics[width=5.2cm,height=4.3cm]{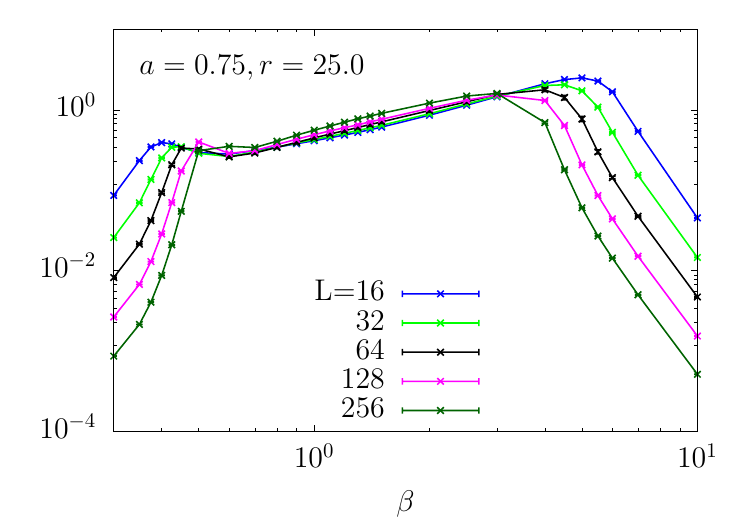}
\includegraphics[width=5.2cm,height=4.3cm]{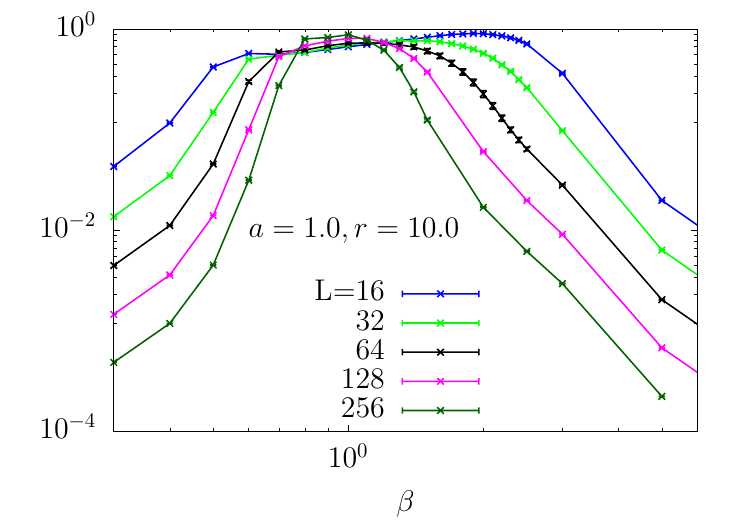}
\end{center}
\vspace{-0.5cm}
\caption{$q=3$ Potts model. 
Rescaled magnetic susceptibility,  $ 0.33^{L/16} \chi(L,\beta)$,  as a function of $\beta$ for $a=0.5$ (a), $a=0.75$ (b), and $a=1.0$ (c).
The value of $r$ is written as a label in the panels and the system sizes are given in the keys. 
}
\label{FigQ3xi}
\end{figure}

In Fig.~\ref{FigQ3xi}, we show the rescaled magnetic susceptibility $0.33^{L/16} \chi(L,\beta)$ for each case as a 
function of $\beta$. The rescaling factor was chosen arbitrarily in order 
to obtain an approximate collapse of the data at $\beta=1$, the critical point.

In the case of uncorrelated disorder, the magnetic susceptibility exhibits a single peak at $\beta=1$.
For $a=0.5$, by contrast, there are two peaks, one located at $\beta \simeq 0.3$ and the other at $\beta \simeq 15$. 
Between these two peaks, the magnetic susceptibility appears to scale with the system size as if the whole region were critical.

A similar result was obtained by Chatelain, who argued that this behaviour is a manifestation of a Griffiths phase \cite{C}. 
We present here an alternative interpretation.
 Indeed, Fig.~\ref{FigQ3xi} shows that the double-peak structure is present 
only for $a=0.5$. For $a=1.0$, although a remnant of the two-peak structure can be observed for the smallest system size, 
$L=16$, a central peak clearly develops as the system size increases. For $a=0.75$, the situation is less clear. We still 
observe a two-peak structure, but both peaks move slowly towards the center, particularly the one located at $\beta<1$.
For this latter peak, we measured the position $\beta_{\rm max}(L)$ of the maximum of the magnetic susceptibility and found
\begin{align}
\beta_{\rm max}(L) = 0.96 \pm 0.06 - 0.73 L^{-0.1},
\end{align}
which is compatible with a single peak located at $\beta=1$ in the thermodynamic limit.

Note that, to determine the peak positions accurately, we performed additional simulations in the vicinity of the maxima and increased 
the number of disorder realisations by a factor of ten for $a=0.75$ and $a=0.5$, the latter of which will be discussed next. These additional data are not shown here.

For $a=0.5$, we also observe that the two peaks move with increasing system size, although very slowly. 
By studying the position of the maximum of the magnetic susceptibility as a function of $L$, we find that the peak located at $\beta<1$ behaves as
\begin{align}
\beta_{\rm max}(L) = 0.36 \pm 0.01 - 0.18L^{-0.23},
\end{align}
indicating a finite limiting value of $\beta_{\rm max}$.

\vspace{0.25cm}

\begin{figure}[!ht]
\centerline{
(a) \hspace{7cm} (b) \hspace{3.5cm}
}
\begin{center}
\includegraphics[width=7.75cm,height=6.5cm]{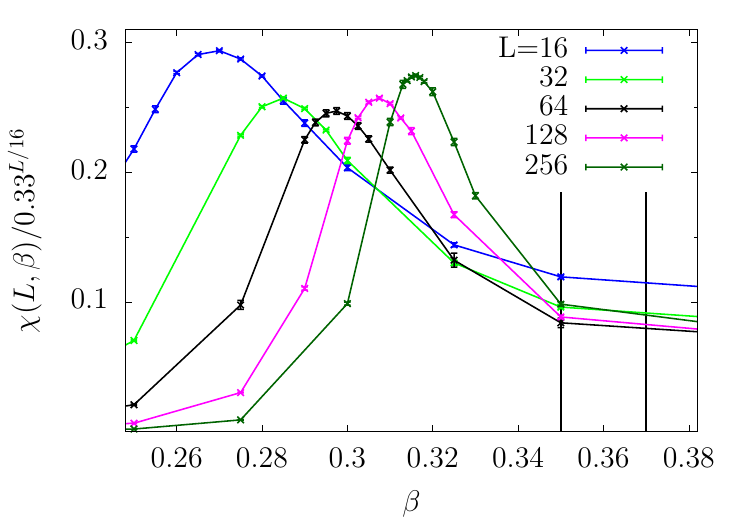}
\includegraphics[width=7.75cm,height=6.5cm]{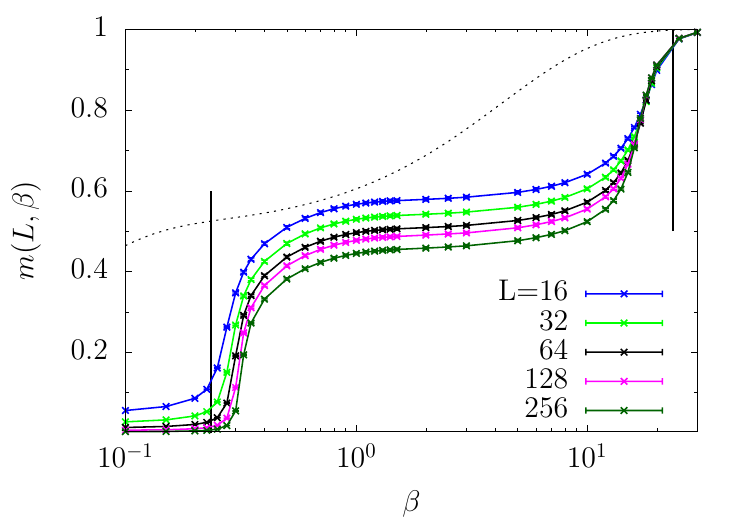}
\end{center}
\vspace{-0.5cm}
\caption{$q=3$ Potts model with $a=0.5$. (a) Scaled linear magnetic susceptibility as a function of the inverse temperature $\beta$ for different system 
sizes. The drift towards higher values of $\beta$  of the position at $\beta_{\rm max}$ of the left peak  is clear. The vertical segments
are the limits of the  interval $\beta_{\rm max}(\infty) = 0.36\pm 0.01$. 
(b) Magnetisation density as a function of $\beta$. The two scales $\beta \sim 0.23$ and $\beta\sim 23$, for the onset of 
strong bonds and activation of weak bonds, respectively, are shown with vertical 
black segments. See the text for a discussion of the dotted curve.
}
\label{FigQ3xi_c}
\end{figure}

The corresponding measurements are shown in Fig.~\ref{FigQ3xi_c}(a). The two vertical black lines indicate the bounds obtained for the extrapolated value of the peak position, namely $\beta_{\rm max}(\infty)=0.36 \pm 0.01$. We therefore conclude that the double-peak structure of the magnetic susceptibility persists only for sufficiently small values of $a$, corresponding to the regime in which the system flows towards an infinite-disorder fixed point.

For large disorder strength, $J_1 \to 0$ while $J_2 = rJ_1$ becomes large. 
For a given spin configuration, two spins are connected by a bond with probability $p_1(\beta)=1-\exp(-\beta J_1)$ 
for a weak bond and $p_2(\beta)=1-\exp(-\beta J_2)$ for a strong bond.

In the infinite-disorder limit, $p_1 \to 0$ and $p_2 \to 1$, and the model reduces to percolation (with no temperature dependence).

For $r=100$, we have $J_2=4.25$ and $J_1=0.0425$. The corresponding temperature scale $\beta \simeq 1/J_2 \simeq 0.2353$ sets the onset of strong-bond formation. 
A second temperature scale occurs at $\beta \simeq 1/J_1 \simeq 23.53$, corresponding to the activation of weak bonds. This behaviour is illustrated in  Fig.~\ref{FigQ3xi_c}(b).

For $\beta \ll 0.2353$, no clusters are formed and the magnetisation is very small. Around $\beta \simeq 0.2353$, strong bonds become activated,\footnote{A bond between spins $\sigma_i$ and $\sigma_j$ is activated with probability $(1-\exp(-\beta J_{ij})) \delta_{\sigma_i,\sigma_j}$.} leading to the formation of large clusters. The magnetisation exhibits a sharp increase and the magnetic susceptibility diverges.

For $0.2353 \ll \beta \ll 23.53$, only strong bonds contribute to cluster formation. Since the fraction of strong bonds is at the bond percolation threshold $p_c=1/2$, these clusters can percolate, and the magnetisation is finite and increases slowly with $\beta$.

For $\beta \gg 23.53$, weak bonds are also activated, allowing all bonds to contribute to percolation, and the magnetisation approaches one.

In the right panel of Fig.~\ref{FigQ3xi_c}, the values $\beta=0.2353$ and $\beta=23.53$ are indicated by vertical black segments. 
We also show, with a dotted line, the quantity 
$(p_1(\beta)+p_2(\beta))/2$. It approaches $0.5$ near the first vertical line and $1.0$ near the second.

Of course, we are still at finite disorder, so thermal fluctuations remain relevant. A bond is activated only if the two spins it connects are in the same state. 
The dotted line therefore provides an upper bound on the number of activated bonds. In this limit, the dotted curve converges to a constant value $1/2$, 
corresponding to the percolation threshold, while the two characteristic inverse-temperature scales separate, with one tending to zero and the other diverging.

Note that similar behavior was documented for the 3D Ising model by Wang {\it et al.}~\cite{WMLW} considering a decay exponent $a \simeq 1$, where correlated disorder 
induced size-dependent double peaks in the susceptibility that merged asymptotically. We find a parallel mechanism in 2D, where the double peak disappears for large sizes in the presence of a finite disorder fixed point (for $a \geq 0.75$).

\section{$q=8$ Potts model}
\label{sec:q8Potts}

Next, we turn to the $q=8$ Potts model. 
This case is considerably more complicated because, in the absence of disorder, the model undergoes a first-order phase transition. 
Consequently, the weak-disorder perturbative approach cannot be applied. Indeed, in the pure system, the correlation length is finite, with $\xi \simeq 24$ on the square lattice for $q=8$ \cite{BW}.
In the absence of disorder, $W(r=1,L)$ is a function of the ratio $L/\xi$. As disorder is introduced, the correlation length $\xi$ increases \cite{CJ,JP}. Therefore, $W(r,L)$ depends on both $L$ and the disorder-dependent correlation length $\xi(r)$, and hence also explicitly on $r$. At present, we do not know how to describe the behaviour of $W(r,L)$ in the weak-disorder regime, where the correlation length remains finite. For this reason, in the following we restrict our analysis to data with $r \geq 2$.

\subsection{Phase diagram and critical points}

\begin{figure}[ht]
\centerline{
(a) \hspace{7cm} (b) \hspace{3.5cm}
}
\begin{center}
\includegraphics[width=7.75cm,height=6.5cm]{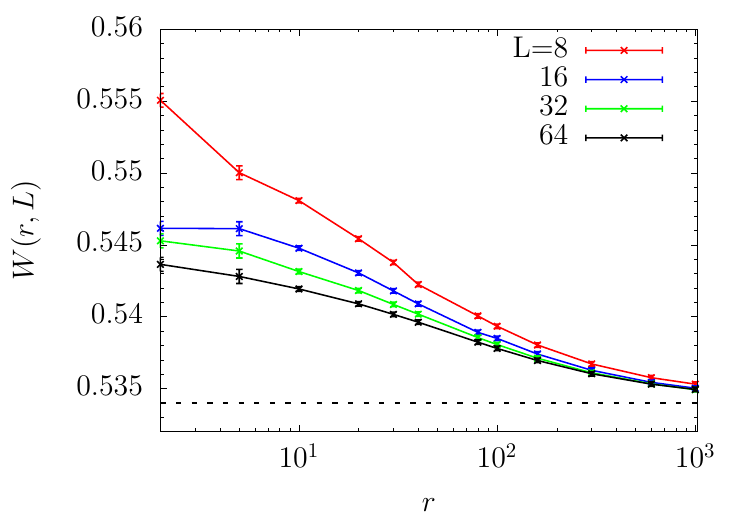}
\includegraphics[width=7.75cm,height=6.5cm]{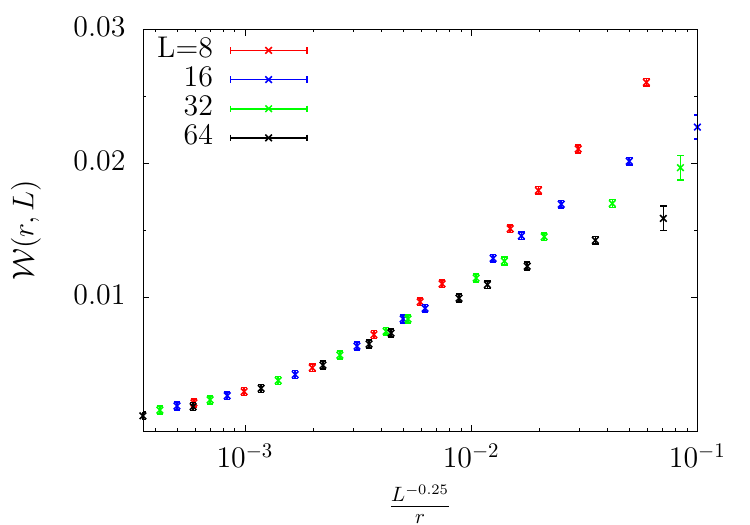}
\end{center}
\vspace{-0.5cm}
\caption{$q=8$ Potts model with $a=0.5$. (a) Wrapping probability as a function of the disorder strength $r$. 
The horizontal dashed line is at $W_1^{a=0.5}=0.5340$.
(b) ${\mathcal W}(r,L)$ defined in Eq.~(\ref{DW}) vs. $L^{-0.25}/r$. Systems sizes and colour code are given in the keys.
}
\label{FigQ8a05}
\end{figure}

We first consider long-range correlated disorder with $a=0.5$.  Again, in Fig.~\ref{FigQ8a05}(a) 
we find that there is no crossing of the wrapping probabilities, 
indicating a flow up towards infinite disorder. 

\vspace{0.25cm}

In Fig.~\ref{FigQ8a05}(b) we plot
\begin{align}
\label{DW}
{\cal W}(r,L) = \left(
\frac{W(r,L)}{W_1^{a=1}(L)} - 1
\right),
\end{align}
vs. $L^{y_p}/r$. This quantity is similar to the previously defined ${\cal W}_2$, see Eq.~(\ref{DW2}). 
We find excellent data collapse in the large-disorder regime for $y_p=-0.25$. The negative value of $y_p$ confirms that the renormalisation-group flow is towards the infinite-disorder fixed point. Note that in this case, each measurement for $r\geq 10$ is averaged over $10^7$ disorder realisations.

\vspace{0.25cm}
\begin{figure}[h!]
\centerline{
(a) \hspace{7cm} (b) \hspace{3.5cm}
}
\begin{center}
\includegraphics[width=7.75cm,height=6.5cm]{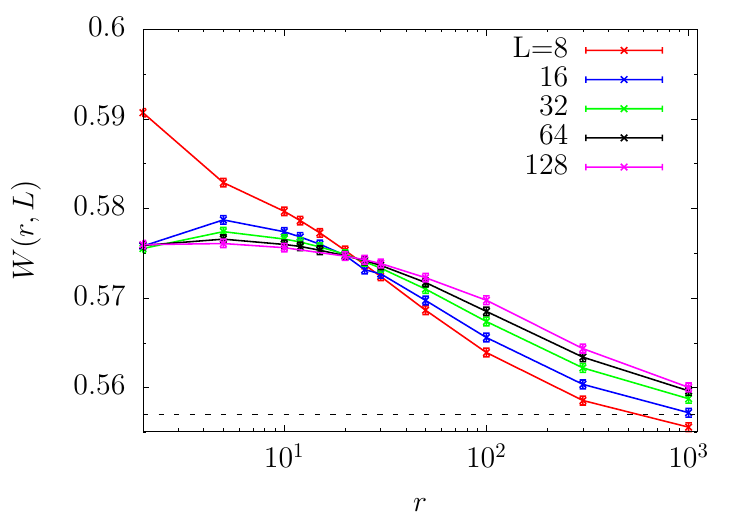}
\includegraphics[width=7.75cm,height=6.5cm]{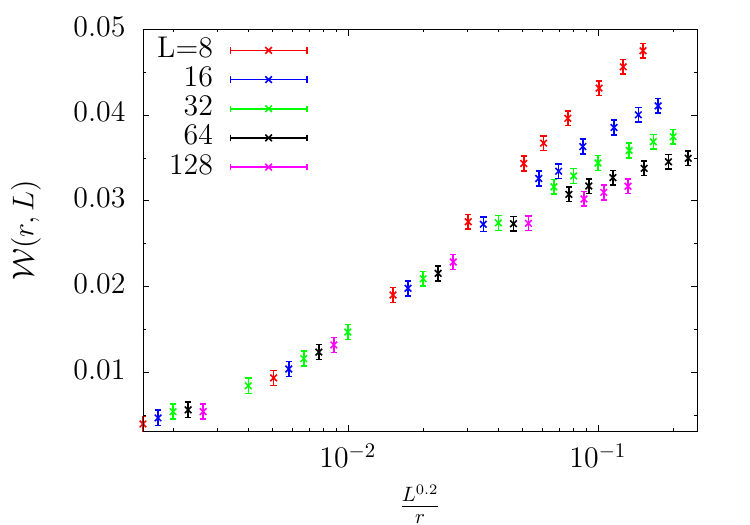}
\end{center}
\vspace{-0.5cm}
\caption{$q=8$ Potts model with $a=0.75$. (a) Wrapping probability as a function of the disorder strength  $r$. 
The horizontal dashed line is at $W_1^{a=0.75}=0.5580$.
(b) ${\mathcal W}(r,L)$ defined in Eq.~(\ref{DW}) vs. $L^{0.2}/r$.
}
\label{FigQ8a075}
\end{figure}

Next, we consider the case $a=0.75$. The data are shown in Fig.~\ref{FigQ8a075}. In Fig.~\ref{FigQ8a075}(a) we plot the wrapping probability $W(r,L)$ as a function of the disorder strength. We observe a crossing of the curves (for $L>8$), but it occurs at a very small disorder strength ($r \simeq 2$) and is again associated with the crossover from a first-order to a second-order phase transition.
A second crossing is observed at $r \simeq 25$. In Fig.~\ref{FigQ8a075}(b), we plot ${\mathcal W}(r,L)$ as a function of 
$L^{0.2}/r$. Since the exponent is positive, this indicates a flow away from the infinite-disorder fixed point. In the following, we therefore take $r=25$ as the location of the fixed point.

\begin{figure}[h!]
\centerline{
(a) \hspace{7cm} (b) \hspace{3.5cm}
}
\begin{center}
\includegraphics[width=7.75cm,height=6.5cm]{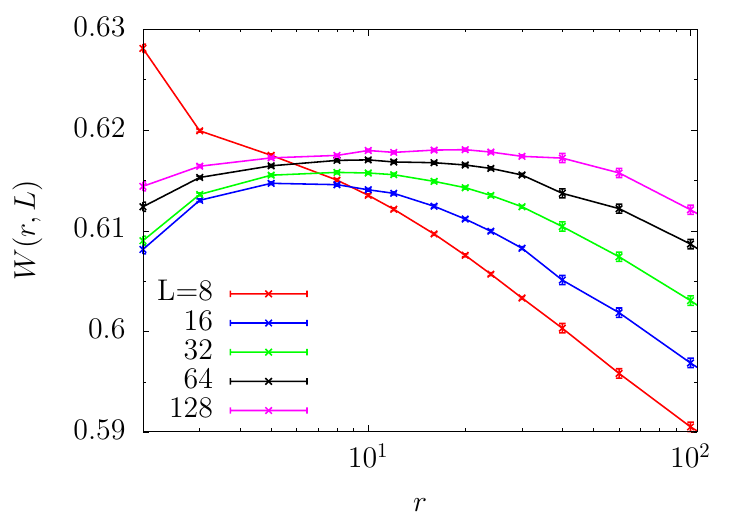}
\includegraphics[width=7.75cm,height=6.5cm]{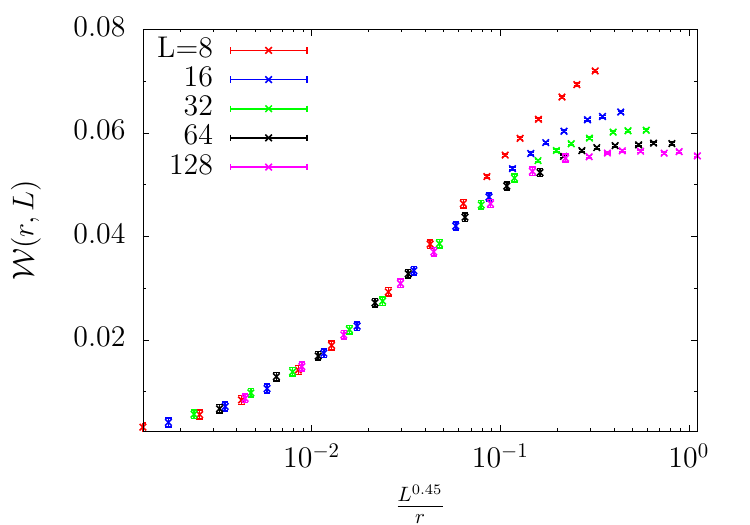}
\end{center}
\vspace{-0.5cm}
\caption{$q=8$ Potts model with $a=1$.
(a) Wrapping probability as a function of the disorder strength $r$. 
(b) ${\mathcal W}(r,L)$ defined in Eq.~(\ref{DW}) vs. 
$L^{0.45}/r$.
}
\label{FigQ8a1}
\end{figure}

In Fig.~\ref{FigQ8a1}(a), we show the wrapping probabilities for $a=1.0$. It is difficult to identify a clear crossing of the curves. 
Nevertheless, Fig.~\ref{FigQ8a1}(b) shows that, for sufficiently large disorder, the wrapping probability exhibits scaling as a function of $L^{0.45}/r$. 
The positive exponent indicates that the infinite-disorder fixed point is repulsive. We therefore expect the existence of a finite-disorder fixed point.

\vspace{0.25cm}

\begin{figure}[ht]
\centerline{
(a) \hspace{7cm} (b) \hspace{3.5cm}
}
\begin{center}
\includegraphics[width=7.75cm,height=6.5cm]{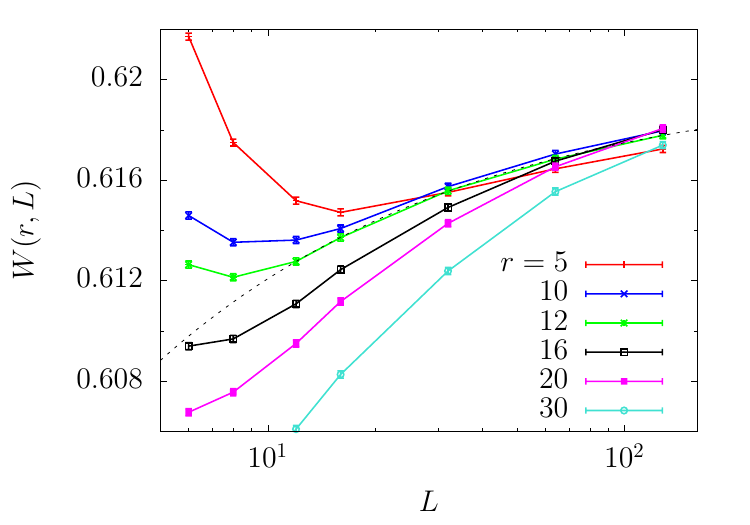}
\includegraphics[width=7.75cm,height=6.5cm]{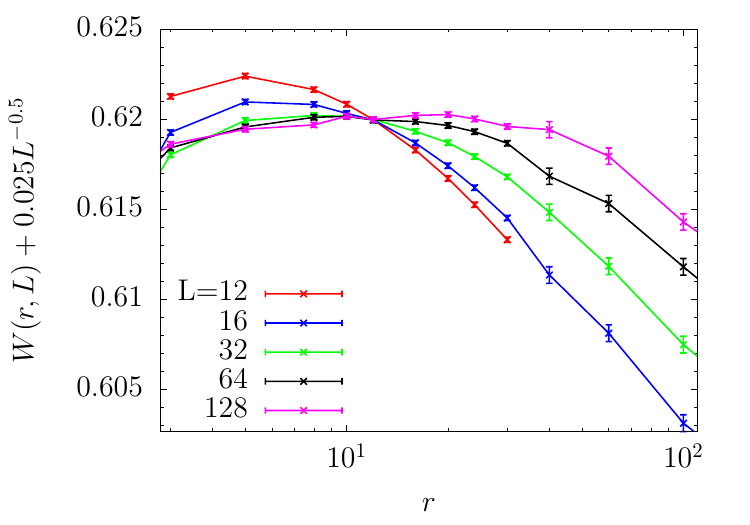}
\end{center}
\vspace{-0.5cm}
\caption{$q=8$ Potts model with $a=1$.
(a) Wrapping probability as a function of $L$ for different values of $r$ given in the key. (b) $W(r,L)+0.025 \, L^{-0.5}$ as a function 
of the disorder strength $r$.
}
\label{FigQ8a1_2}
\end{figure}

In Fig.~\ref{FigQ8a1_2}(a), we plot $W(r,L)$ as a function of $L$ for various values of $r$. For $r$ in the range $10$--$20$, 
$W(r,L)$ converges to a common value as $L$ increases. For weaker disorder, much larger system sizes are required to reach the same limiting value. 
We illustrate this behavior with the case $r=5$. A similar behavior is observed for stronger disorder, as illustrated by the case $r=30$. 
For both values of $r$, $W(r,L)$ also appears to converge to the same limiting value, but only at much larger system sizes.
Note that we increased the statistics to $10^7$ samples and included additional data for $L=6$ and $L=12$.

Close to a fixed point, we expect $W(r,L)$ to contain two size-dependent terms.\footnote{We introduce $\sigma=(J_1-J_2)/2$, which measures the strength of the disorder, with $\sigma_c$ denoting its value at the fixed point.}
\begin{align}
W(r,L) = W_0 + \alpha_1 (\sigma^2 - \sigma^2_c) L^{-|y|} + \alpha_2 L^{-\omega} \; ,
\end{align}
the first term corresponds to the flow near the fixed point, with a negative exponent $-|y|$ because the fixed point is attractive, while the second term corresponds to irrelevant operators.
A complete fit of these data is too complicated. In practice, it is more convenient to consider the simpler fitting form
\begin{align}
\label{FW}
W(r,L) = W_0 + \alpha_0 L^{-\omega_0} \; ,
\end{align}
which reduces to the previous expression when $\sigma^2 = \sigma_c^2$. 
We obtain good fits for all values of $r \in [12,24]$. In each case, we find $\omega_0 \simeq 0.5$, while $\alpha_0$ depends almost linearly on $\sigma^2$. 
Its absolute value is minimal near $r=12$ (note that $\alpha_0$ is always negative). 
The simplest interpretation is therefore that $y \simeq \omega \simeq 0.5$ and that the fixed point is located close to $r=12$. 
Figure~\ref{FigQ8a1_2}(a) shows the best fit of Eq.~(\ref{FW}) for $r=12$ with $L \geq 10$. 
The fitted values are $\alpha_0=-0.025$ and $\omega_0=0.5$. In Fig.~\ref{FigQ8a1_2}(b), we plot $W(r,L)+0.025 \, L^{-0.5}$ and observe a crossing at $r=12$.

\vspace{0.25cm}

\subsection{Magnetic susceptibility}

\begin{figure}[!ht]
\centerline{
\hspace{-0.8cm} (a) \hspace{4.5cm} (b) \hspace{4.5cm} (c) \hspace{2cm}
}
\begin{center}
\includegraphics[width=5.2cm,height=4.2cm]{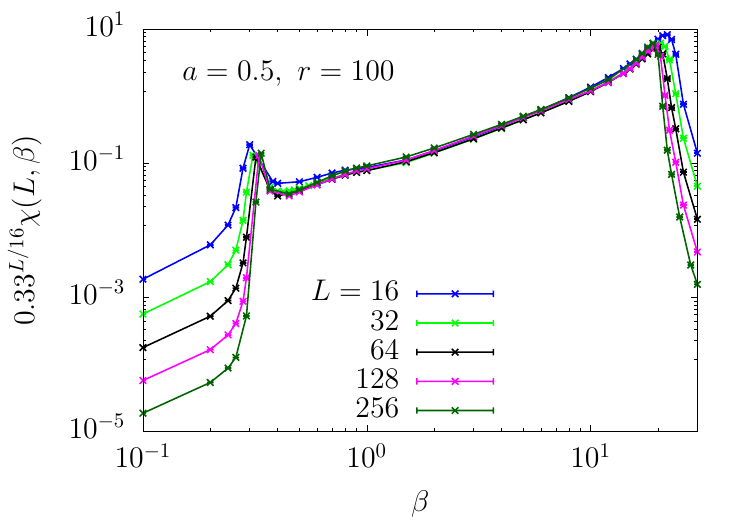}
\includegraphics[width=5.2cm,height=4.2cm]{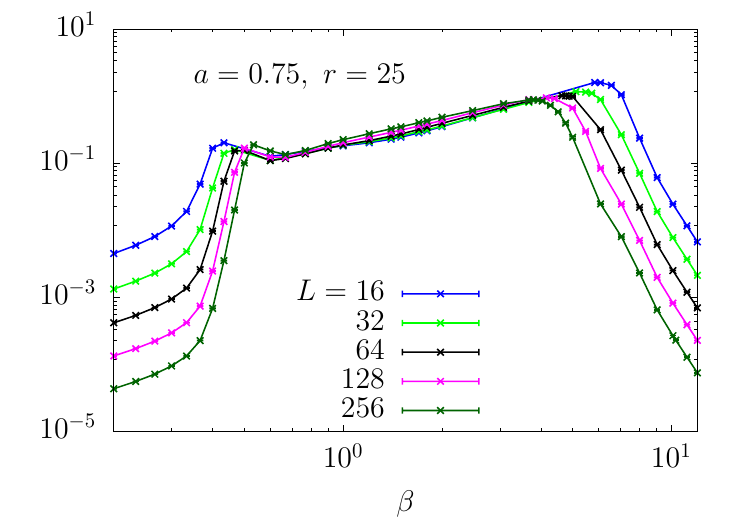}
\includegraphics[width=5.2cm,height=4.2cm]{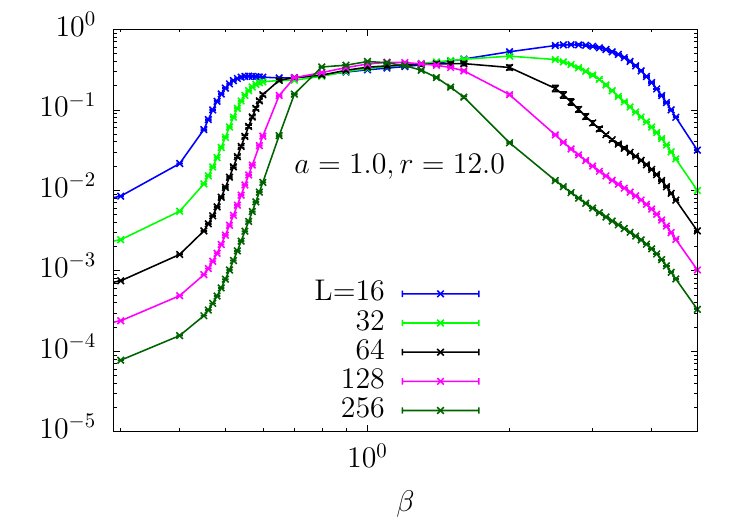}
\end{center}
\vspace{-0.5cm}
\caption{$q=8$ Potts model.
Rescaled magnetic susceptibility, $0.33^{L/16} \chi(L,\beta)$, as a function of $\beta$ for $a=0.5$ (a), $a=0.75$ (b), and $a=1.0$ (c).
The corresponding values of $r$ are indicated as labels within the panels.
}
\label{FigQ8xi}
\end{figure}

Next, we present the results of our simulations for the magnetic susceptibility. As for the case $q=3$, we consider three pairs of $a$ and $r$:
(i) $a=0.5$ and $r=100$; (ii) $a=0.75$ and $r=25$; and (iii) $a=1.0$ and $r=12$.
The result of our measurements are shown in Fig.~\ref{FigQ8xi}. It contains the rescaled magnetic susceptibility, $0.33^{L/16} \chi(L,\beta)$ as a 
function of $\beta$
where again, $\beta=1$ corresponds to the critical inverse temperature. The results are similar to the ones found for the
$q=3$ Potts model. 

\begin{figure}[!ht]
\begin{center}
\includegraphics[width=7.75cm,height=6.5cm]{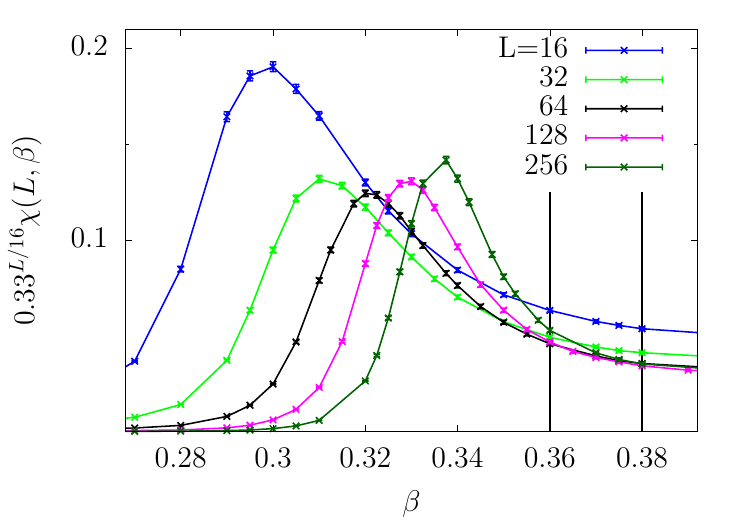}
\end{center}
\vspace{-0.5cm}
\caption{Position of the left peak in the $q=8$ Potts model with $a=0.5$ and $r=100$.
The vertical black segments are located at  $\beta_{\rm max}(L\to \infty) = 0.37 \pm 0.01$.
}
\label{FigQ8xi_c}
\end{figure}

For $a=0.5$, we clearly observe two peaks, and they move slowly as one increase the linear system size $L$. 
A fit of the position of the maximum of the left peak 
converges towards a finite value
\begin{align}
\beta_{\rm max}(L) = 0.37 \pm 0.01 - 0.14 L^{-0.26} \; .
\end{align}
Figure~\ref{FigQ8xi_c} shows the positions of the peaks as well as the bounds for their  extrapolated value shown as two vertical black lines. 

Thus, for $a=0.5$, the two peaks structure will survive in the infinite size limit. We expect that it is for the same reasons as for $q=3$, the left peak 
corresponds to the activation of strong bonds while the right peak is for the activation of weak bonds. 

For $a=0.75$, the positions of the maximum of the left peak increase near linearly with the size. An extrapolation gives a value for $\beta_{\rm max}$ larger than one.
Thus, for this case one expects a single peak at $\beta=1$, as for the $q=3$ model. For $a=1$, the existence of a single peak is quite apparent. 

\section{Hyper-scaling} 
\label{sec:hyperscaling}

\begin{figure}[!ht]
\centerline{
(a) \hspace{7cm} (b) \hspace{3.5cm}
}
\begin{center}
\includegraphics[width=7.75cm,height=6.5cm]{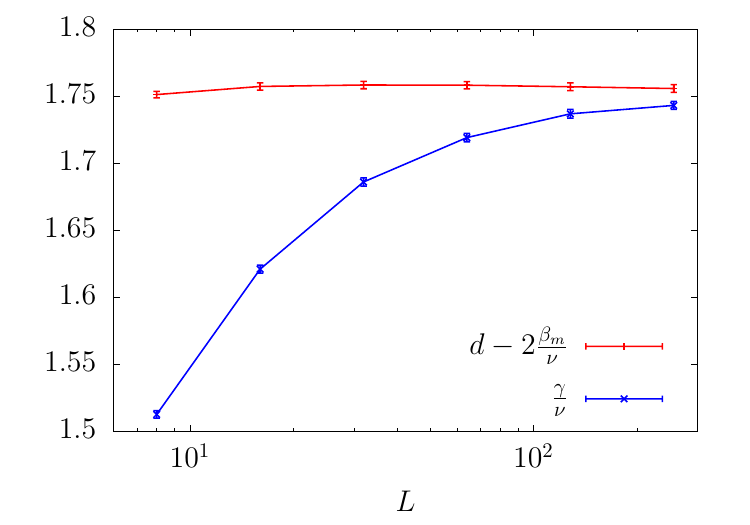}
\includegraphics[width=7.75cm,height=6.5cm]{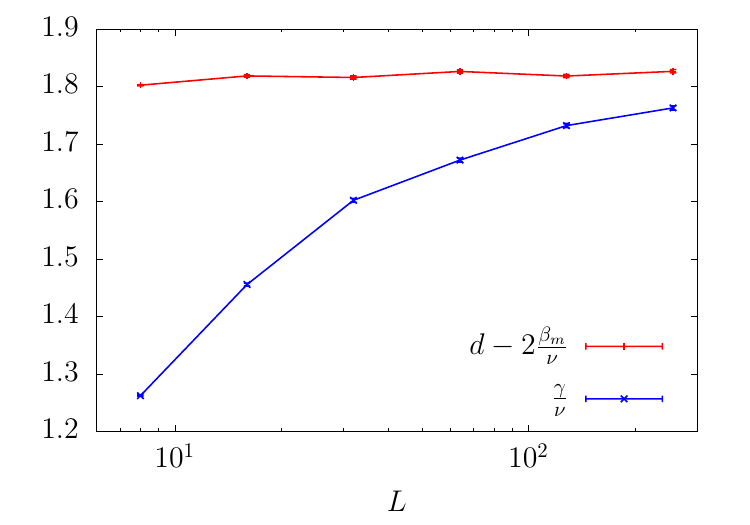}
\end{center}
\vspace{-0.5cm}
\caption{Effective exponents $d-2 \, \beta_m / \nu$ (red) and $\gamma/\nu$ (blue) vs. $L$. 
(a) $q=3$ Potts model with $a=1.0$ and $r=10$ 
and (b) $q=8$ Potts model with $a=0.5$ and $r=100$.  
}
\label{FigExp3}
\end{figure}

In this section, we present our results on hyperscaling. It was argued by Chatelain \cite{C,C2} that hyperscaling is violated in the Potts model with long-range interactions.
This was inferred, in particular, from measurements of the exponent associated with the magnetic susceptibility:
\begin{align}
\chi(L,\beta)  = L^2 \left( \overline{\langle m^2\rangle - \langle m\rangle ^2} \right)  \simeq L^{\frac{\gamma}{\nu}}  \; .
\end{align}
The hyperscaling relation states that
\begin{equation}
\frac{\gamma}{\nu} = 2 - 2\, \frac{\beta_m}{\nu}
\; , 
\end{equation}
where $\beta_m/\nu$ is the scaling dimension associated with the magnetisation:
\begin{align}
\overline{\langle m\rangle } \simeq L^{-\frac{\beta_m}{\nu}} 
\;  .
\end{align}
We denote by $\beta_m$ the exponent associated with the magnetisation to avoid any confusion with the inverse temperature $\beta$.
Since we consider only the critical case $\beta=1$ throughout the remainder of the paper, we omit the explicit dependence on $\beta$ from the notation for simplicity.
In Fig.~\ref{FigExp3}, we show the effective exponents obtained from our measurements, 
\begin{align}
(\frac{\gamma}{\nu})^{({\rm eff})}(L)=\dfrac{\log{ \Big(\dfrac{\chi(2L)}{\chi(L)} } \Big) } {\log{2}}  \; , 
\qquad \quad
 \; (\frac{\beta_m}{\nu})^{({\rm eff})}(L)=-\dfrac{\log{ \Big( \dfrac{\overline{\langle m\rangle }(2L)}{\overline{\langle m\rangle }(L)}} \Big) } {\log{2}}  \; .
\end{align}
Let us first discuss the data for the $q=3$ Potts model with $a=1$ and $r=10$, shown in (a). 
Similar results are obtained for the case shown in (b), corresponding to $q=8$, $a=0.5$, and $r=100$. The magnetic scaling dimension $\beta_m/\nu$ takes the value $\simeq 0.12$, in agreement with previous results \cite{CPS1}. We observe only small finite-size corrections for this quantity, and consequently also for $2-2\, \beta_m/\nu \simeq 1.76$. In contrast, $\gamma/\nu$ exhibits strong finite-size corrections as a function of the system size $L$. For small system sizes, it takes an effective value of approximately $1.6$.
In Figs.~\ref{FigQ3xi} and \ref{FigQ8xi}, the magnetic susceptibility was rescaled by $0.33^{L/16}$. Consequently, at each doubling of the system size $L$, it was multiplied by a factor $0.33 \simeq 1/2^{1.6}$. However, these figures already show that, for the largest system sizes, the data collapse at the critical point $\beta=1$ is not satisfactory, despite the logarithmic scale used for the $y$-axis. In the large-system-size limit, the effective exponent $\gamma/\nu$ converges to the same value as $2 - 2\, \beta_m/\nu$, indicating that the hyperscaling relation is satisfied. Figure~\ref{FigExp3} illustrates this behavior for two representative cases, and we have verified that it holds for all values of $a$ and $q$ that we investigated. In particular, this behavior is observed both for the small value $a=0.5$, for which the fixed point is located at infinity, and for the larger value $a=1$, for which the fixed point is located at a finite value. 

\section{Conclusion}
\label{sec:conclusions}

In this work, we investigated the critical behavior and phase diagram of the two-di\-men\-sional $q$-state Potts model ($q=3$ and $q=8$) 
subject to long-range, power-law-correlated quenched disorder. By analysing the wrapping probabilities of Fortuin--Kasteleyn clusters, 
we refined the phase diagram proposed in Ref.~\cite{CPS1}. Furthermore, using measurements of the magnetic susceptibility and 
magnetisation density, we resolved the physical nature of the double-peak structure and tested the hyperscaling relation.

Our main findings are summarised as follows.
\begin{itemize}
\item \textbf{Fixed Point Structure:} From our wrapping probability analysis, we established that for $a \ge 0.75$, both the $q=3$ and $q=8$ Potts models possess 
an attractive finite-disorder fixed point corresponding to a second-order phase transition. Conversely, for $a = 0.5$, the systems flow 
toward an infinite-disorder fixed point characterised by percolation-like scaling behavior.
    
\item \textbf{Origin of the Double Peak:} While disorder-averaged magnetic susceptibility curves exhibit a double-peak structure 
as a function of inverse temperature $\beta$, finite-size scaling reveals two distinct physical regimes:
\begin{itemize}
   \item For $a \ge 0.75$, the double peak is an intermediate finite-size artefact. As system size $L$ increases, the peaks merge into a single 
   well-defined transition at $\beta = 1$, mirroring the finite-size behavior documented by Wang \textit{et al.}~\cite{WMLW} for the 3D Ising model with $a \approx 1$.
   \item For $a = 0.5$, the double-peak structure persists in the thermodynamic limit. Rather than indicating an extended Griffiths phase as previously suggested~\cite{C}, we interpret this persistence as a two-scale activation process where strong and weak bond energy scales decouple, reducing the system to bond percolation at criticality.
    \end{itemize}
    
    \item \textbf{Validity of Hyperscaling:} By taking into account strong finite-size corrections in the ratio $\gamma/\nu$, we confirmed that the hyperscaling relation $\gamma/\nu = 2 - 2\beta_m/\nu$ remains satisfied across all investigated values of $a$ and $q$, clarifying previous reports of hyperscaling violation~\cite{C, C2}.
\end{itemize}

Taken together, our findings clarify the interplay between long-range correlated disorder and strong fluctuations, offering a comprehensive picture of 
criticality in low-dimensional disordered Potts models.

\vspace{0.5cm}

\noindent
{\bf \large
Acknowledgements}

This study was financed in part by the Coordenação de Aperfeiçoamento de Pessoal de Nível Superior -- Brasil (CAPES) -- Finance Code 001.


\end{document}